\documentstyle{elsart}

\begin{document}
\begin{frontmatter}
\title{Star Formation Histories in the Local Group}
\author{Thomas M. Brown}
\address{Space Telescope Science Institute\\ 3700 San Martin Drive, Baltimore,
MD 21218, USA}

\begin{abstract}
Deep color magnitude diagrams extending to the main sequence provide
the most direct measure of the detailed star formation history in a
stellar population. With large investments of observing time, HST can
obtain such data for populations out to 1 Mpc, but its field of view
is extremely small in comparison to the size of Local Group
galaxies. This limitation severely constrains our understanding of
galaxy formation.  For example, the largest galaxy in the Local Group,
Andromeda, offers an ideal laboratory for studying the formation of
large spiral galaxies, but the galaxy shows substructure on a variety
of scales, presumably due to its violent merger history. Within its
remaining lifetime, HST can only sample a few sight-lines through this
complex galaxy. In contrast, a wide field imager could provide a map
of Andromeda's halo, outer disk, and tidal streams, revealing the
spatially-dependent star formation history in each structure. The same
data would enable many secondary studies, such as the age variation in
Andromeda's globular cluster system, gigantic samples of variable
stars, and microlensing tracers of the galaxy's dark matter distribution.
\end{abstract}
\end{frontmatter}

\section{Historical Background}
For more than 50 years \cite{S53}, a color-magnitude diagram (CMD)
reaching the main sequence (MS) has been the most direct tool for measuring
the star formation history in a stellar population.  Much of that work
focused on the ages of simple stellar populations, such as star
clusters, within our own Galaxy.  By the late 1980s, the leap in 
sensitivity and photometric precision offered by the widespread
use of CCDs, coupled with advances in stellar evolution models,
allowed accurate estimates for the ages of Galactic globular clusters
\cite{V88}.  In the 1990s, the launch of the Hubble Space Telescope
(HST) and the subsequent installation of the Wide Field Planetary
Camera 2 enabled the measurement of star formation histories in more
complicated populations throughout the Local Group.  Beyond the
closest satellites of the Milky Way, these studies focused on the
brighter and younger populations in dwarf galaxies \cite{DP02}.

\begin{figure}
\includegraphics{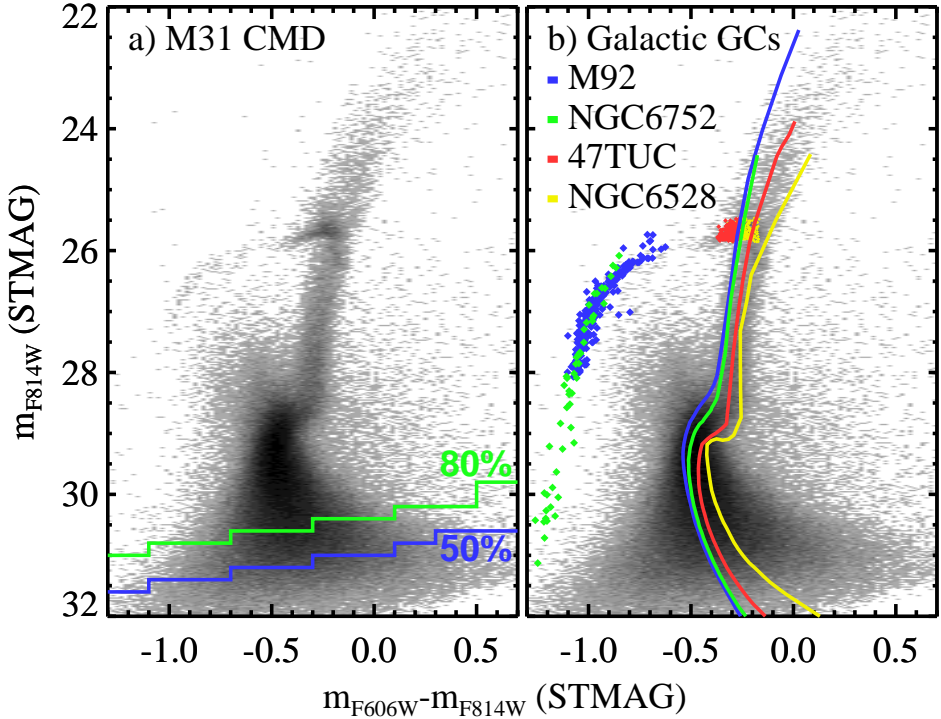}
\vspace{3.8in}
\caption{{\it Panel a:} The CMD of the M31 halo, as observed by ACS
in the F606W (broad $V$) and F814W ($I$) bands \cite{B03}, with
completeness limits marked.  {\it Panel b:} Comparison 
to the ACS observations of 4 Galactic globular clusters
in the same bands.  The relative MS turnoff luminosities
demonstrate empirically that the M31 halo ranges from metal-rich
stars at intermediate ages to metal-poor stars at old ages. 
Comparison to theoretical isochrones calibrated in the ACS bands also shows
that over 50\% of the M31 halo stars are at higher metallicity ([Fe/H]$> -0.5$)
and younger ages (6--11~Gyr) than the stars in our own halo
\cite{B03}.}
\end{figure} 

\section{The Advanced Camera for Surveys}
In 2001, the Advanced Camera for Surveys (ACS) was installed on HST,
and its dramatic improvements in sensitivity, field of view, and
sampling enabled dating of the oldest populations out to the edge of
the Local Group.  Extremely deep ACS imaging of Andromeda (M31) showed
that its halo is surprisingly younger than our own \cite{B03}, with
over half of the halo younger than 11~Gyr, and with 30\% of
the population at intermediate ages of 6--8~Gyr (Figure 1).  The data
from such long exposures are also useful for studying transient
phenomena (e.g., microlensing, supernovae) and variable stars
\cite{B04}.  Soon two more deep fields will be obtained
in M31's outer disk and giant tidal stream.
Because the ACS field is extremely small compared to the size
of Local Group galaxies (e.g., the M31 halo is several degrees
across), HST can only produce pencil-beam samples in each major
component of a nearby galaxy.  M31's complex substructure, seen
in maps of its bright giant stars \cite{F02}, cannot be sampled
at this depth within the remaining HST lifetime.

\begin{figure}
\includegraphics{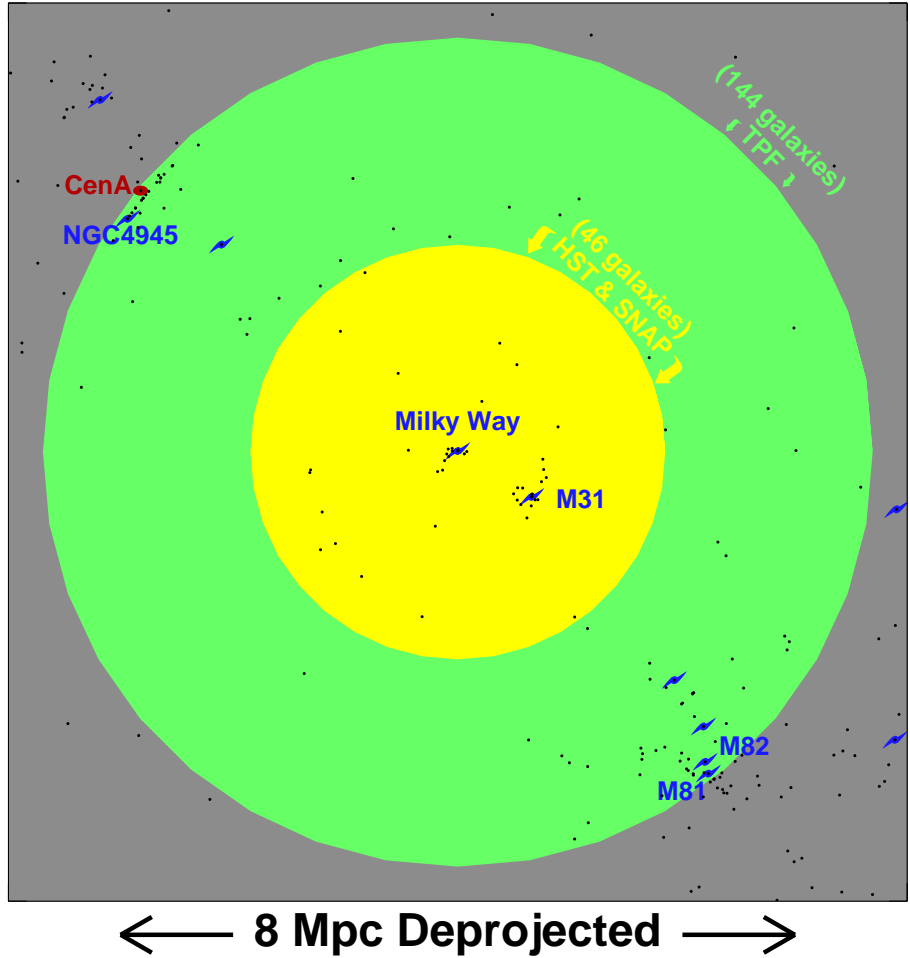}
\vspace{5.0in}
\caption{A deprojected view of the galaxies in the local Universe
(black points), assuming recent estimates of their distance and distribution
\cite{K04}, showing the volume that can be surveyed down to an old (12~Gyr)
MS turnoff with telescopes from various missions.  Giant
galaxies are marked by spiral and elliptical symbols.  HST
and SNAP can survey the Local Group, while TPF
can reach nearly 4 Mpc, assuming 200 hours of
observations per field in the F606W and F814W bands.}
\vspace{0.15in}
\end{figure}

\section{The Future: Wide Field Imaging from Space}
There are two near-term avenues for major advances in the study of
star formation histories utilizing upcoming optical space missions
(Figure 2): the Supernova Acceleration Probe (SNAP) and the
Terrestrial Planet Finder (TPF).  The James Webb Space
Telescope will also make progress, but its infrared bands do not
provide as good a temperature lever for MS stars.

As an optical telescope with the sensitivity and sampling of
HST, but with a much larger field of view, SNAP could map the star
formation history in the outer disk and halo of large galaxies like
M31 and M33.  The observing program might consist of two stages:
an exploratory stage, where shallow tiles across the galaxy
provide secondary age diagnostics on the red giant branch and
horizontal branch, and a deep followup stage, providing direct age
diagnostics via MS photometry, sampling at a $\approx$20\%
fill factor.  Current plans envision 0.1$^{\prime\prime}$ pixels;
0.05$^{\prime\prime}$ is better suited to surveys of the M31
outer disk, while larger pixels would be restrict this work to the
halo outskirts.  Its focal plane design (with many fixed filters) does
not provide an efficient way to obtain two-band photometry over large
fields, but the additional bands would provide tighter population
constraints.

The optical coronograph version of the TPF, with its $4 \times 6$
meter aperture, will provide high-contrast, high-resolution imaging
over tiny fields of view as its prime mission.  However, a secondary
instrument with a wide field of view could obtain MS photometry in
galaxies nearly 4 Mpc away.  Currently, we are limited to the study of
the two giant galaxies in the Local Group (our own and Andromeda) plus
a handful of smaller galaxies, but the volume of space that could be
surveyed with TPF would include the Cen-A Group and the M81-M82 Group
(Figure 2), greatly expanding the sample of galaxies.


\begin{thebibliography}{99}

\bibitem{S53}
A. Sandage, 
The color-magnitude diagram for the globular cluster M 3, 
{\em AJ} {\bf 58} (1953) 62--75.

\bibitem{V88}
D.A. VandenBerg,
Ages of Galactic globular clusters,
in: J.E. Grindlay and A.G.D. Philip, eds., 
{\em Proceedings of the 126th symposium of the IAU}
(Reidel, Dordrecht, 1988) 107--120.

\bibitem{DP02}
R.C. Dohm-Palmer, E.D. Skillman, M. Mateo, A. Saha, A. Dolphin, E.
Tolstoy, J.S. Gallagher, and A.A. Cole,
Deep Hubble Space Telescope Imaging of Sextans A. I. The Spatially Resolved 
Recent Star Formation History,
{\em AJ} {\bf 123} (2002) 813-831. 

\bibitem{B03}
T.M. Brown, H.C. Ferguson, E. Smith, R.A. Kimble, A.V. Sweigart, A. Renzini,
R.M. Rich, and D.A. VandenBerg,
Evidence for a Significant Intermediate-Age Population in the M31 Halo from 
Main Sequence Photometry,
{\em ApJ} {\bf 592} (2003) L17-L20.

\bibitem{B04}
T.M. Brown, H.C. Ferguson, E. Smith, R.A. Kimble, A.V. Sweigart, A. Renzini,
and R.M. Rich, 
RR Lyrae Stars in the Andromeda Halo from Deep Imaging with the Advanced 
Camera for Surveys,
{\em AJ} {\bf 127} (2004) 2738-2752.

\bibitem{F02}
A.M.N. Ferguson, M.J. Irwin, R.A. Ibata, G.F. Lewis, and N.R. Tanvir,
Stellar Substructure in the Halo and Outer Disk of M31,
{\em AJ} {\bf 124} (2002) 1452-1463.

\bibitem{K04}
I.D. Karachentsev, V.E. Karachentseva, W.K. Huchtmeier, and D.I. Makarov,
A Catalog of Neighboring Galaxies,
{\em AJ} {\bf 127} (2004) 2031-2068.

\end{thebibliography}
\end{document}